\documentclass[journal]{IEEEtran}

\usepackage{graphicx}
\usepackage{booktabs}
\usepackage{subfigure}
\usepackage{overpic}
\usepackage{amsmath}
\usepackage{flushend}
\usepackage{amssymb}
\usepackage{multirow}
\usepackage{makecell}
\usepackage{array}
\usepackage{stfloats}
\usepackage{algorithm,amsfonts}
\usepackage{epsfig, hyperref}
\usepackage{soul}
\usepackage{color, xcolor}

\hypersetup{
colorlinks=true,
linkcolor=blue,
urlcolor=blue,
citecolor=blue,
}

\begin{document}

\title{Hybrid Convolutional and Attention Network for Hyperspectral Image Denoising}

\author{Shuai Hu, Feng Gao, \textit{Member, IEEE}, Xiaowei Zhou, Junyu Dong, \textit{Member, IEEE}, \\ and Qian Du, \textit{Fellow, IEEE}
\thanks{This work was supported in part by the National Key
Research and Development Program of China under Grant 2022ZD0117202
and in part by the Natural Science Foundation of Qingdao under Grant
23-2-1-222-ZYYD-JCH. (Corresponding author: Xiaowei Zhou.)

Shuai Hu, Feng Gao, Xiaowei Zhou, and Junyu Dong are with the School of Computer Science and Technology, Ocean University of China, Qingdao 266100, China.

Qian Du is with the Department of Electrical and Computer Engineering, Mississippi State University, Starkville, MS 39762 USA.}}

\markboth{}
{Shell }

\maketitle

\begin{abstract}

Hyperspectral image (HSI) denoising is critical for  the effective analysis and interpretation of hyperspectral data. However, simultaneously modeling global and local features is rarely explored to enhance HSI denoising. In this letter, we propose a hybrid convolution and attention network (HCANet), which leverages both the strengths of convolution neural networks (CNNs) and Transformers. To enhance the modeling of both global and local features, we have devised a convolution and attention fusion module aimed at capturing long-range dependencies and neighborhood spectral correlations. Furthermore, to improve multi-scale information aggregation, we design a multi-scale feed-forward network to enhance denoising performance by extracting features at different scales. Experimental results on mainstream HSI datasets demonstrate the rationality and effectiveness of the proposed HCANet. The proposed model is effective in removing various types of complex noise. Our codes are available at \url{https://github.com/summitgao/HCANet}.

\end{abstract}


\IEEEpeerreviewmaketitle

\section{Introduction}

\IEEEPARstart{H}{yperspectral} imaging is a powerful technique that allows the acquisition of rich spectral information from an object or a scene. Compared with RGB data, hyperspectral image (HSI) captures fine-grained spectral information. Hence, HSIs have been extensively used in many practical applications, such as unmixing \cite{unmixing} and ground object classification \cite{classification}. However, HSI is often afflicted by inevitable mixed noise generated during the sensor imaging process, which is caused by insufficient exposure time and reflected energy. These noise may degrade the image quality and hinder the performance of subsequent analysis and interpretation. Eliminating these noise could improve the accuracy of ground object detection and classification. Therefore, HSI denoising is a critical and indispensable technique in the preprocessing stage of many remote sensing applications. 

Motivated by the spatial and spectral properties of HSI, traditional HSI denoising methods exploit the optimization schemes with priors, such as low rankness \cite{lowrank1}, total variation \cite{yuan12tgrs}, non-local similarities \cite{ngmeet}, and spatial-spectral correlation \cite{6909773}. Although these methods have achieved appreciable performance, they commonly depends on the degree of similarity between the handcrafted priors and the real-world noise model. In recent years, convolutional neural networks (CNNs) \cite{CNN1} have provided new ideas for HSI denoising, demonstrating notable performance advancements. Maffei et al. \cite{maffei20tgrs} proposed a CNN-based HSI denoising model by taking the noise-level map as input to train the network. Wang et al. \cite{wang23jstars} proposed a convolutional network based on united Octave and attention mechanism for HSI denoising. Pan et al. \cite{pan23jstars} presented a progressively multiscale information aggregation network to remove the noise in HSI. These CNN-based methods use convolution kernels for local feature modeling. 

More recently, with the emergence of Vision Transformer (ViT) \cite{vit}, Transformer-based methods have achieved significant success in various computer vision tasks. Existing Transformer-based image denoising methods have achieved great success through learning the global contextual information. However, if local features are considered and exploited effectively, the HSI denoising performance may improve further. So, it is important to take account of both local and global information by combining CNN and Transformers to enhance denoising performance.

It is commonly non-trivial to build an effective Transformer and CNN hybrid model for HSI denoising, due to the following two challenges: \emph{1) The optimal hybrid architecture for local and global feature modeling still remains an open question.} Convolutional kernels capture local features, which means losing the long-distance information interaction. The combination of convolution and attention could offer a viable solution. \emph{2) The single-scale feature aggregation of the feed-forward network (FFN) in Transformer is limited.} Some methods use depth-wise convolution to improve local feature aggregation in FFN. However, due to the larger number of channels in the hidden layer, single-scale token aggregation can hardly exploit rich channel representations. 

\begin{figure*}[ht]
\centering
\includegraphics [width=6in]{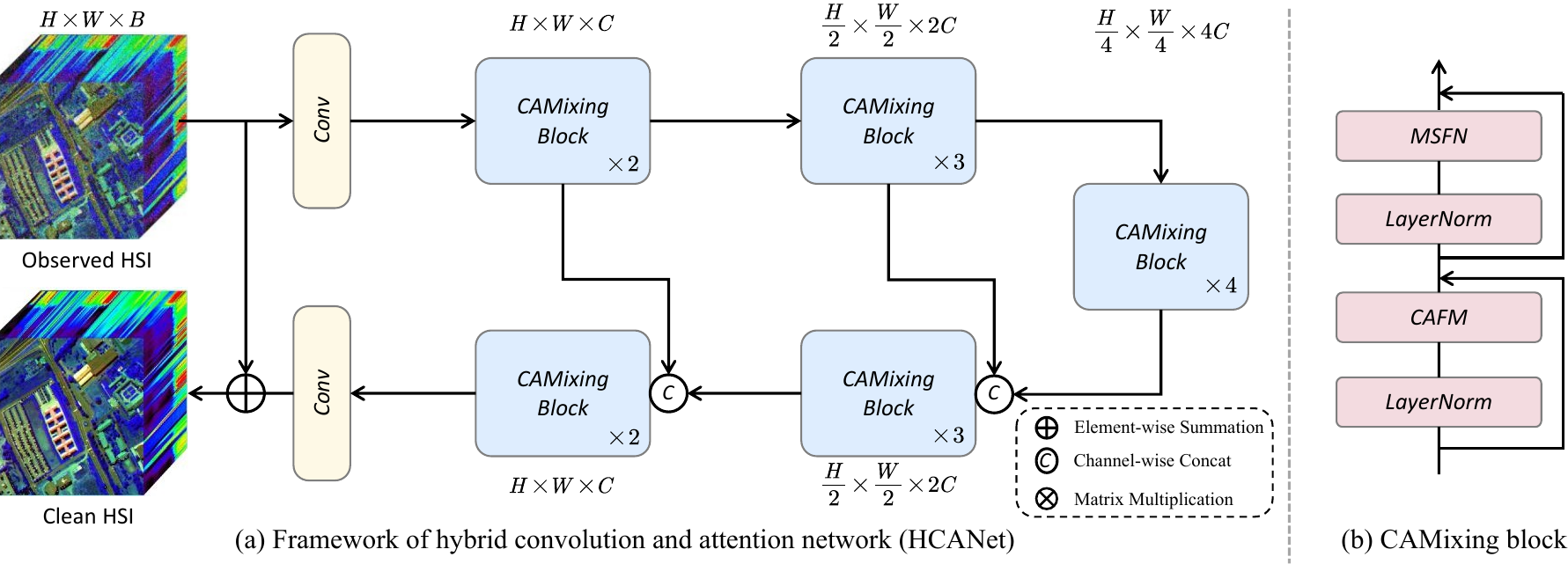}
\caption{Illustration of our proposed hybrid convolution and attention network (HCANet) for HSI denoising. (a) Framework of HCANet. (b) Inner structure of CAMixing block.}
\label{fig_framework}
\end{figure*}

To solve the aforementioned two challenges, we proposed a \underline{H}ybrid \underline{C}onvolution and \underline{A}ttention  \underline{Net}work (HCANet) for HSI denoising, which simultaneously exploits both the global contextual information and local features, as illustrated in Fig. \ref{fig_framework}. Specifically, to enhance the modeling of both global and local features, we have devised a convolution and attention fusion module (CAFM) aimed at capturing long-range dependencies and neighborhood spectral correlations. Furthermore, to improve multi-scale information aggregation in FFN, we design a multi-scale feed-forward network (MSFN) to enhance denoising performance by extracting features at different scales. Three parallel dilated convolutions with different strides are used in MSFN. By conducting experiments on two real-world datasets, we validate that our proposed HCANet is superior to other state-of-the-art competitors.

The contributions of this letter can be summarized as follows:

\begin{itemize}

\item The promising yet challenging problem of global and local feature modeling for HSI denoising is explored. To the best of our knowledge, this is the first work to combine convolution and attention mechanism for the HSI denoising task. 
    
\item We propose multi-scale feed-forward network, which seamlessly extract feature at different scales, and effectively suppress the noise from multiple scales.
    
\item Extensive experiments are conducted on two benchmark datasets, which demonstrates the rationality and effectiveness of the proposed HCANet.  As a side contribution, we have released our codes to benefit other researchers.
\end{itemize}

\section{Methodology}

In this section, we present the Hybrid Convolution and Attention Network (HCANet) for hyperspectral image denoising. As shown in Fig. \ref{fig_framework}, the main structure of the model is a U-shaped network with several Convolution Attention Mixing (CAMixing) blocks. Each CAMixing block includes two parts: convolution-attention fusion module (CAFM) and multi-scale feed-forward network (MSFN). For HSI, 3D convolutions capture spatial and spectral features comprehensively, but increase parameters. To manage complexity, we use 2D convolutions for channel adjustment, effectively exploiting HSI features.

For a noisy hyperspectral image $\mathbf{I} \in \mathbb{R}^{H \times W \times B}$, where $H \times W$ denote the spatial resolution and $B$ denotes the channel dimension, our HCANet first uses $3\times3\times3$ convolution to obtain low-level feature. Then, we use a U-shaped network with several CAMixing blocks and skip connections to obtain the noise residual map $\mathbf{I}_N \in \mathbb{R}^{H \times W \times B}$, which has the same shape as the input noisy image. The reconstructed clean HSI can be expressed as: $\hat{\mathbf{I}}=\mathrm{I}+\mathrm{I}_N$. Finally, the reconstruction loss with a global gradient regularizer is used to train the HCANet.  

HSI denoising aims to reconstruct the corresponding clean HSI from a noisy HSI. Fusing global and local features in HSI is important to enhance the denoising task. Thus, the CAFM is designed in the CAMixing block. To further utilize the contextual cues in the feed-forward network for HSI denoising, the MSFN is designed in the CAMixing block. Next, we present the details of the CAFM and MSFN.

\subsection{Convolution and Attention Fusion Module}

\begin{figure}[]
\centering
\includegraphics [width=3in]{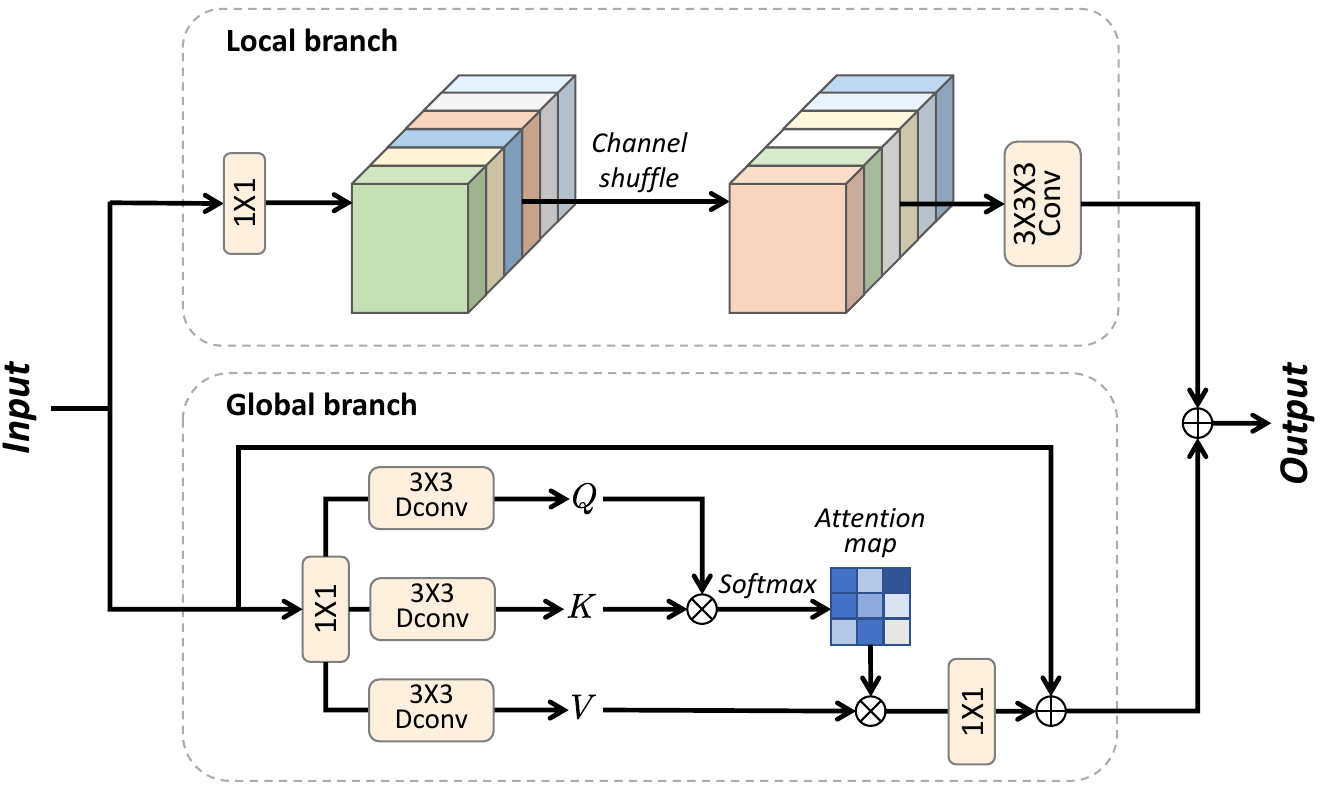}
\caption{Illustration of the proposed convolution and attention fusion module (CAFM). It consists of local and global branches. In the local branch, convolution and channel shuffling are employed for local feature extraction. In the global branch, the attention mechanism is used to model long-range feature dependencies.}
\label{fig_cafm}
\end{figure}

Convolutional operations, limited by their local nature and restricted sensory field, are insufficient in modeling global features. In contrast, the Transformer, enabled by the attention mechanism, excels in extracting global features and capturing long-range dependencies. Convolution and attention are complementary to each other to model both global and local features. Inspired by this, we design a convolution and attention fusion module (CAFM), as shown in Fig. \ref{fig_cafm}. We employ a self-attention mechanism in the global branch to capture broader hyperspectral data information, while a local branch focuses on extracting local features for comprehensive denoising.

The proposed CAFM consists of local and global branches. In the local branch, to enhance cross-channel interaction and promote information integration, we first use $1\times1$ convolution to adjust the channel dimensions. Following this, a channel shuffling operation is performed to further mix and blend the channel information. Channel shuffle partitions the input tensor along the channel dimension into groups, wherein depth-wise separable convolution is employed within each group to induce channel shuffling. Subsequently, the resulting output tensors from each group are concatenated along the channel dimension to generate a novel output tensor. Subsequently, we utilize a $3\times3\times3$ convolution to extract features. The local branch can be formulated as:

\begin{equation}
\setlength{\abovedisplayskip}{1ex}
    \textbf{F}_{\rm conv} = W_{3\times3\times3}({\rm CS}(W_{1\times1}(\mathbf{Y}))),
\setlength{\belowdisplayskip}{1ex}
\end{equation}
where $\textbf{F}_{\rm conv}$ is the output of local branch, $W_{1\times1}$ denotes $1\times1$ convolution, $W_{3\times3\times3}$ denotes $3\times3\times3$ convolution, CS represents channel shuffle operation, and $\mathbf{Y}$ is the input feature.

In the global branch, we first generates query(\textbf{Q}), key(\textbf{K}) and value(\textbf{V}) via $1\times1$ convolution and $3\times3$ depth-wise convolution, yielding three tensors with the shape of $\hat{H}\times\hat{W}\times\hat{C}$. Next, \textbf{Q} is reshaped to $\hat{\textbf{Q}}\in\mathbb{R}^{\hat{H}\hat{W}\times\hat{C}}$, and $K$ is reshaped to $\hat{\textbf{K}}\in\mathbb{R}^{\hat{C}\times\hat{H}\hat{W}}$. Then, we compute the attention map $\mathbf{A}\in\mathbb{R}^{\hat{C}\times \hat{C}}$ via the interaction of $\hat{\textbf{Q}}$ and $\hat{\textbf{K}}$. The computational burden is reduced instead of computing the huge regular attention map of size $\mathbb{R}^{\hat{H}\hat{W}\times\hat{H}\hat{W}}$. The output $\textrm{$\textbf{F}_{\rm att}$}$ of global branch is defined as:
\begin{equation}
  \textrm{$\textbf{F}_{\rm att}$} = W_{1\times1}\textrm{Attention}\left(\hat{\textbf{Q}}, \hat{\textbf{K}}, \hat{\textbf{V}}\right) + \textbf{Y},
\end{equation}
\begin{equation}
  \textrm{Attention}\left(\hat{\textbf{Q}}, \hat{\textbf{K}}, \hat{\textbf{V}}\right) = \hat{\textbf{V}} \textrm{Softmax$\left( \hat{\textbf{K}} \hat{\textbf{Q}}/\alpha \right)$},  
\end{equation}
where $\alpha$ is a learnable scaling parameter to control the magnitude of matrix multiplication of $\hat{\mathbf{K}}$ and $\hat{\mathbf{Q}}$ before applying the softmax function. 

Finally, the output of the CAFM module calculation is computed as:
\begin{equation}
    \textbf{F}_{\rm out} = \textbf{F}_{\rm att} + \textbf{F}_{\rm conv}.
\end{equation}

\subsection{Multi-Scale Feed-Forward Network}

The original FFN in ViT is composed of two linear layers for single-scale feature aggregation. However, the information contained in single-scale feature aggregation of FFN is limited. To enhance the non-linear feature transformation, we propose a multi-scale feed-forward network (MSFN). After each CAMixing block, the outputs of CAFM are fed into the MSFN to aggregate multi-scale features and enhance the non-linear information transformation. Previous studies have revealed the efficacy of incorporating multi-scale information in image denoising tasks \cite{pan23jstars}.

\begin{figure}[]
\centering
\includegraphics [width=3in]{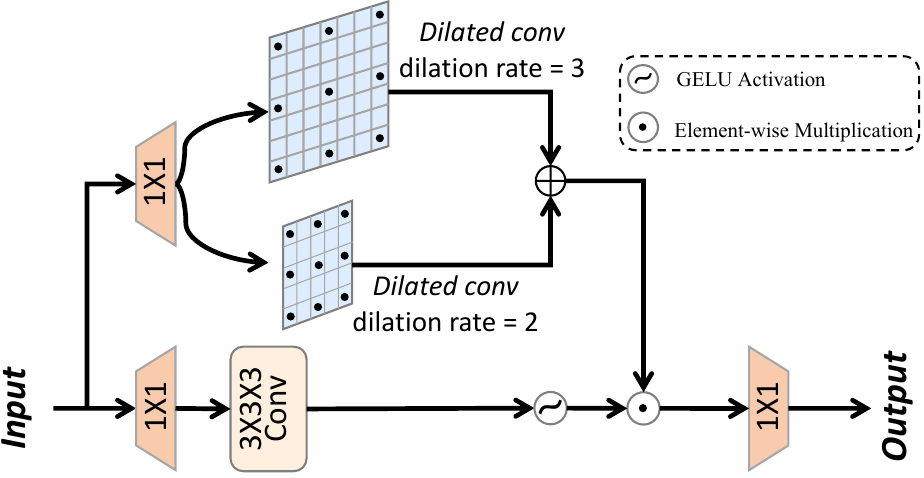}
\caption{Illustration of the multi-scale feed-forward network (MSFN).}
\label{fig_msfn}
\end{figure}

Details of the MSFN are illustrated in Fig. \ref{fig_msfn}. Two $1\times1$ convolutions are used to expand the feature channels with expanding ratio $\gamma=2$.
The input features are handled in two parallel paths, and gating mechanism is introduced to enhance the non-linear transformation via element-wise product of features from both paths. In the lower path, depth-wise convolution is used for feature extraction. In the upper path, multi-scale dilated convolutions are employed for multi-scale feature extraction. Two $3\times3$ dilated convolutions with dilation rates of 2 and 3 are used. 

Given an input tensor $\mathbf{X}\in\mathbb{R}^{\hat{H}\times \hat{W} \times \hat{C}}$,  MSFN is formulated as:

\begin{equation}
\begin{split}
\textrm{Gating}(\textbf{X}) = {\phi}(W_{3\times3\times3} W_{1\times1}(\textbf{X})) ~~~~~~~~~~~~~~~~~~~~\\
~~~~ \odot (W^2_{3\times3} W_{1\times1}(\textbf{X})+W^3_{3\times3} W_{1\times1}(\textbf{X})),
\end{split}
\end{equation}

\begin{equation}
\textrm{$\textbf{X}_{out}$} = W_{1\times1}\textrm{Gating}(\textbf{X}),
\end{equation}
where $\odot$ denotes element-wise multiplication, $\phi$ represents the GELU non-linearity, $W^2_{3\times3}$ denotes $3\times3$ dilated convolution with dilation rate of 2 and  $W^3_{3\times3}$ denotes $3\times3$ dilated convolution with dilation rate of 3. MSFN offers a distinct role compared to CAFM, focusing on enriching features with contextual information.

\subsection{Loss Function}

We use the $L_1$ loss function as our optimization objective for training the network, and the reconstruction loss is:
\begin{equation}
    \mathcal{L}_\textrm{rec} = \Vert \hat{\mathbf{I}} - \mathbf{I} \Vert_1,
\end{equation}
$\hat{\mathbf{I}}$ represents the estimated noise-free Hyperspectral Image (HSI), while $\mathbf{I}$ denotes the noisy HSI. 

To enhance the quality of denoising and curtail redundancy, we incorporate a global gradient regularizer to constrain $\hat{\mathbf{I}}$,

\begin{equation}
\begin{split}
	\mathcal{L}_{grad} = 
	\Vert \nabla_h\hat{\mathbf{I}} - \nabla_{h}\mathbf{I} \Vert _2^2 + \Vert \nabla_v\hat{\mathbf{I}} - \nabla_{v}\mathbf{I} \Vert _2^2 + \\
	\Vert \nabla_s\hat{\mathbf{I}} - \nabla_{s}\mathbf{I} \Vert _2^2 ~~~~~~~~~~~~~~~~~~~~~~~~
\end{split}
\end{equation}
where $\nabla_h$, $\nabla_v$, and $\nabla_s$ respectively represent the gradient operator applied along the horizontal, vertical, and spectral axes. Finally, the total loss function is as follows:
\begin{equation}
	\mathcal{L} = \mathcal{L}_{rec} + \lambda \mathcal{L}_{grad},
\end{equation}
where $\lambda$ is the weight parameter governing $\mathcal{L}_{grad}$. Based on empirical evidence, we set $\lambda$ as 0.01 to maintain a balance between the various loss terms.

\begin{table}[tbp]
\caption{Quantitative evaluation of the proposed HCANet and other comparing methods in HSI image denoising task with four noise magnitudes ($\sigma = 30,50,70$ and $blind$).}
\resizebox{\columnwidth}{!}{
\begin{tabular}{c|c|ccccccc}
\toprule
Case  & Index & Noisy & LRMR  & BM4D  & LRTV  & Restormer  & MAFNet & Ours \\ \midrule
& PSNR  & 17.26  & 31.81 & 39.71 & 36.92 & 42.19 & {\color[HTML]{4472C4} 43.95} & {\color[HTML]{FF0000} 45.51} \\  
 & SSIM  & 0.108  & 0.678 & 0.973 & 0.93 &  0.971 & {\color[HTML]{4472C4}0.974} & {\color[HTML]{FF0000} 0.981} \\ 
\multirow{-3}{*}{30} & SAM & 0.698 & 0.182 & 0.079 & 0.089 & 0.05 & {\color[HTML]{4472C4} 0.031} & {\color[HTML]{FF0000} 0.028} \\ \midrule
 & PSNR  & 15.43  & 30.04 & 37.22 & 35.99 & 41.51 & {\color[HTML]{4472C4} 42.13}  & {\color[HTML]{FF0000} 44.54} \\ 
 & SSIM  & 0.049 & 0.634 & 0.856 & 0.901 & 0.952 & {\color[HTML]{4472C4} 0.967}  & {\color[HTML]{FF0000} 0.978} \\  
\multirow{-3}{*}{50}  & SAM & {\color[HTML]{374151} 0.885} & 0.222 & 0.166 & 0.121 & 0.047 & {\color[HTML]{4472C4} 0.034} & {\color[HTML]{FF0000} 0.033} \\ \midrule
 & PSNR  & {\color[HTML]{374151} 11.45} & 25.89 & 34.09 & 33.88 & 39.31 & {\color[HTML]{4472C4} 41.05} & {\color[HTML]{FF0000} 43.27} \\ 
 & SSIM  & 0.03 & 0.565 & 0.781 & 0.858 & 0.947 & {\color[HTML]{4472C4} 0.951}  & {\color[HTML]{FF0000} 0.968} \\  
\multirow{-3}{*}{70} & SAM & 1.01 & 0.275 & 0.189 & 0.155 & 0.041 &{\color[HTML]{FF0000}  0.036} & {\color[HTML]{4472C4} 0.037} \\ \midrule
 & PSNR  & 14.85  & 30.07 & 37.1  & 37.23 & 40.73 & {\color[HTML]{4472C4} 42.21}  & {\color[HTML]{FF0000} 43.97} \\
& SSIM  & 0.056 & 0.648 & 0.867 & 0.924 & 0.95 & {\color[HTML]{4472C4} 0.962}  & {\color[HTML]{FF0000} 0.978} \\  
\multirow{-3}{*}{blind} & SAM   & {\color[HTML]{374151} 0.857} & 0.149 & 0.086 & 0.114 & 0.043  & {\color[HTML]{FF0000} 0.032} & {\color[HTML]{4472C4} 0.034} \\ \bottomrule
\end{tabular}

}
\label{tab1}
\end{table}

\section{Experimental Results and Analysis}\label{s3}

\subsection{Experiment Setup}

\textbf{Benchmark datasets.} To verify the denoising performance of our model on hyperspectral images, we trained our model on ICVL dataset and evaluated the trained model on Pavia dataset. In the ICVL dataset, a total of 31 spectral bands were utilized to collect images with a resolution of $1392\times 1300$. To facilitate the training process, the data were randomly cropped and transformed into cube data with a shape of $128\times128\times31$. We augmented the training dataset with rotation and scaling techniques to improve the model's robustness, which resulted in a dataset with 20, 000 new samples. To test the denoising effect of our model on real remote sensing images, we conducted experiments on Pavia dataset.

\textbf{Noise setting.} During the testing phase, we conducted experiments under two settings to demonstrate the effectiveness and generalizability of our proposed model. For the first setting, we tested our model with varying magnitudes of Gaussian noise from $\sigma=30$ to $\sigma=70$ and blind Gaussian noise (random noise magnitude). For the second setting, we evaluated its robustness against common complex noise types found in hyperspectral data obtained from real spaceborne sensors, such as Gaussian, impulse, and deadline noise. We defined five types of complex noise set:

\begin{enumerate}[]

\item \textit{Case 1: Gaussian noise of varying magnitudes in all spectral channels, with a randomly chosen standard deviation from 30 to 70.}

\item \textit{Case 2: Gaussian + Stripe noise.} In addition to Gaussian noise, we randomly add strip noise to spectral bands by polluting 5\% $\sim$ 15\% of columns with strips.

\item \textit{Case 3: Gaussian + Deadline noise.} On the basis of Case 1, we added the deadline noise in the third of the spectral bands. 5\% to 15\% of columns are conflicted with deadlines in each band.

\item \textit{Case 4: Gaussian + Impulse noise.} On the basis of Case 1. Approximately one third of the spectral bands were chosen at random to increase the intensity of impulse noise by a range of 30\% to 70\%.

\item \textit{Case 5: Mixture noise.} Similar to the above cases, each spectral band is affected by \textit{non-i.i.d Gaussian noise} in Case 1. Additionally, each band is randomly affected by a combination of three other types of noise.

\end{enumerate}

\textbf{Baseline methods and Implementation Details.} We compared HCANet with five state-of-the-art methods including model-driven methods and deep learning-based methods. For model-driven methods, we consider BM4D \cite{maggioni13tip}, as well as low-rank methods such as LRMR \cite{zhang14tgrs} and LRTV \cite{he16tgrs}. In terms of deep learning-based methods, we choose the well-known methods, Restormer \cite{Zamir22cvpr} and MAFNet \cite{pan23jstars} for comparison. We used three evaluation metrics including peak signal-to-noise ratio (PSNR), structure similarity (SSIM), and spectral angle mapper (SAM) to quantify the denoising performance. Larger PSNR and SSIM values indicate better denoising results, while smaller SAM values indicate better denoising performance. We trained the models with an initial learning rate of $10^{-4}$ and the learning rate decreased gradually over training epochs. The HCANet was optimized by the Adam optimizer and it was trained for 100 epochs on Gaussian noise and 150 epochs on complex noise. We conducted all the experiments under PyTorch framework on a machine with an NVIDIA GTX 2080Ti GPU, Intel Xeon E5 CPU and 32GB RAM. 

\begin{table}[tbp]
\caption{Quantitative evaluation of the proposed HCANet and other comparing methods in HSI image denoising task with five complex noise cases.}
\resizebox{\columnwidth}{!}{
\begin{tabular}{c|c|ccccccc}
\toprule
Case  & Index & Noisy  & LRMR  & BM4D  & LRTV  & Restormer & MAFNet  & Ours \\ \midrule
& PSNR  & 17.79 & 31.91 & 35.62 & 36.89 & 42.03 & {\color[HTML]{4472C4} 43.48}  & {\color[HTML]{FF0000} 44.11} \\  
& SSIM  & 0.158 & 0.706 & 0.884 & 0.896 &  0.970 & {\color[HTML]{4472C4}0.972} & {\color[HTML]{FF0000} 0.979} \\ 
\multirow{-3}{*}{Case1} & SAM & 0.799 & 0.215 & 0.146 & 0.12  &  0.044 & {\color[HTML]{4472C4}0.037} & {\color[HTML]{FF0000} 0.033} \\ \midrule
& PSNR  & 17.41  & 31.07 & 33.94 & 35.64 &  41.13 & {\color[HTML]{4472C4}42.43}  & {\color[HTML]{FF0000} 43.58} \\
& SSIM  & 0.193 & 0.697 & 0.829 & 0.882 &  0.965 & {\color[HTML]{4472C4}0.967} & {\color[HTML]{FF0000} 0.978} \\ 
\multirow{-3}{*}{Case2} & SAM & 0.808 & 0.238 & 0.153 & 0.182 & 0.05 & {\color[HTML]{4472C4} 0.039} & {\color[HTML]{FF0000} 0.036} \\ \midrule
& PSNR  & 17.42 & 30.04 & 33.77 & 33.82 &  40.16  & {\color[HTML]{4472C4}41.74} & {\color[HTML]{FF0000} 41.79} \\
& SSIM  & 0.15 & 0.74  & 0.866 & 0.871 & {\color[HTML]{4472C4}0.961} &  0.958 & {\color[HTML]{FF0000} 0.973} \\ 
\multirow{-3}{*}{Case3} & SAM & 0.882 & 0.242 & 0.179 & 0.147 &  0.052 & {\color[HTML]{FF0000}0.035} & {\color[HTML]{4472C4} 0.041} \\ \midrule
& PSNR  & 15.12 & 29.34 & 32.61 & 33.25 &  36.89 & {\color[HTML]{4472C4} 37.71} & {\color[HTML]{FF0000} 39.99} \\ 
& SSIM  & 0.126 & 0.703 & 0.887 & 0.816 & 0.932 & {\color[HTML]{4472C4} 0.945} & {\color[HTML]{FF0000} 0.964} \\
\multirow{-3}{*}{Case4} & SAM & 0.891 & 0.286 & 0.193 & 0.174 &  0.083 & {\color[HTML]{4472C4}0.077} & {\color[HTML]{FF0000} 0.056} \\ \midrule
& PSNR  & 14.15 & 26.9  & 31.02 & 30.91 &  36.15 & {\color[HTML]{4472C4} 37.27} & {\color[HTML]{FF0000} 40.57} \\  
& SSIM  & 0.107 & 0.601 & 0.711 & 0.743 & 0.926 &  {\color[HTML]{FF0000}0.978} & {\color[HTML]{4472C4} 0.963} \\
\multirow{-3}{*}{Case5} & SAM & 0.912  & 0.305 & 0.22  & 0.198 &  0.079 & {\color[HTML]{4472C4}0.067}  & {\color[HTML]{FF0000} 0.052} \\ \bottomrule
\end{tabular}
}
\label{tab2}
\end{table}

\subsection{Experimental Analysis}
A comprehensive quantitative comparison of our proposed HCANet and other benchmarking methods is shown in Table~\ref{tab1} and \ref{tab2}. In the two tables,  values highlighted by \textcolor{red}{Red} indicate the best while values highlighted by the \textcolor{blue}{Blue} indicate the second best performance.

\begin{figure}[htb]
\centering
\includegraphics [width=3.4in]{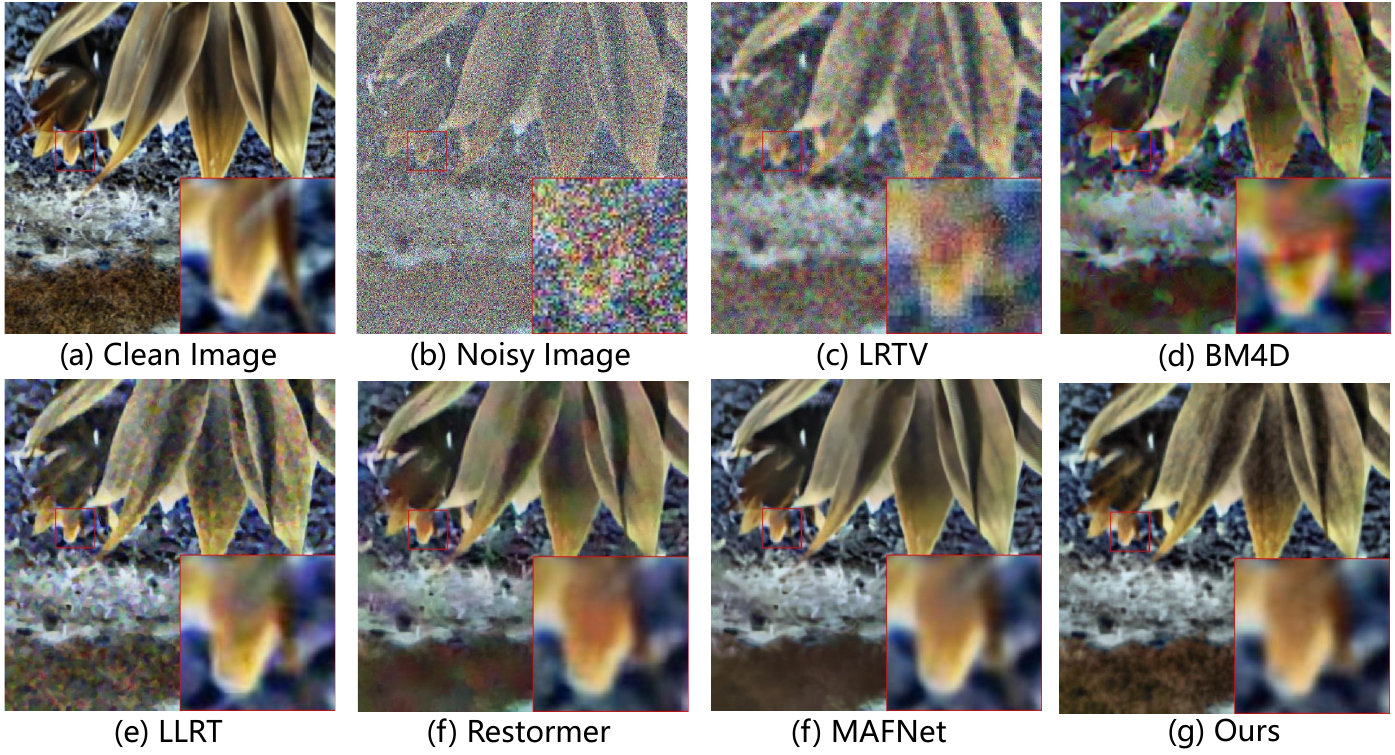}
\caption{Gaussian noise removal results under noise level $\sigma=50$ on ICVL dataset with bands (17,20,30).}
\label{result1}
\end{figure}

\begin{figure}[htb]
\centering
\includegraphics [width=3.4in]{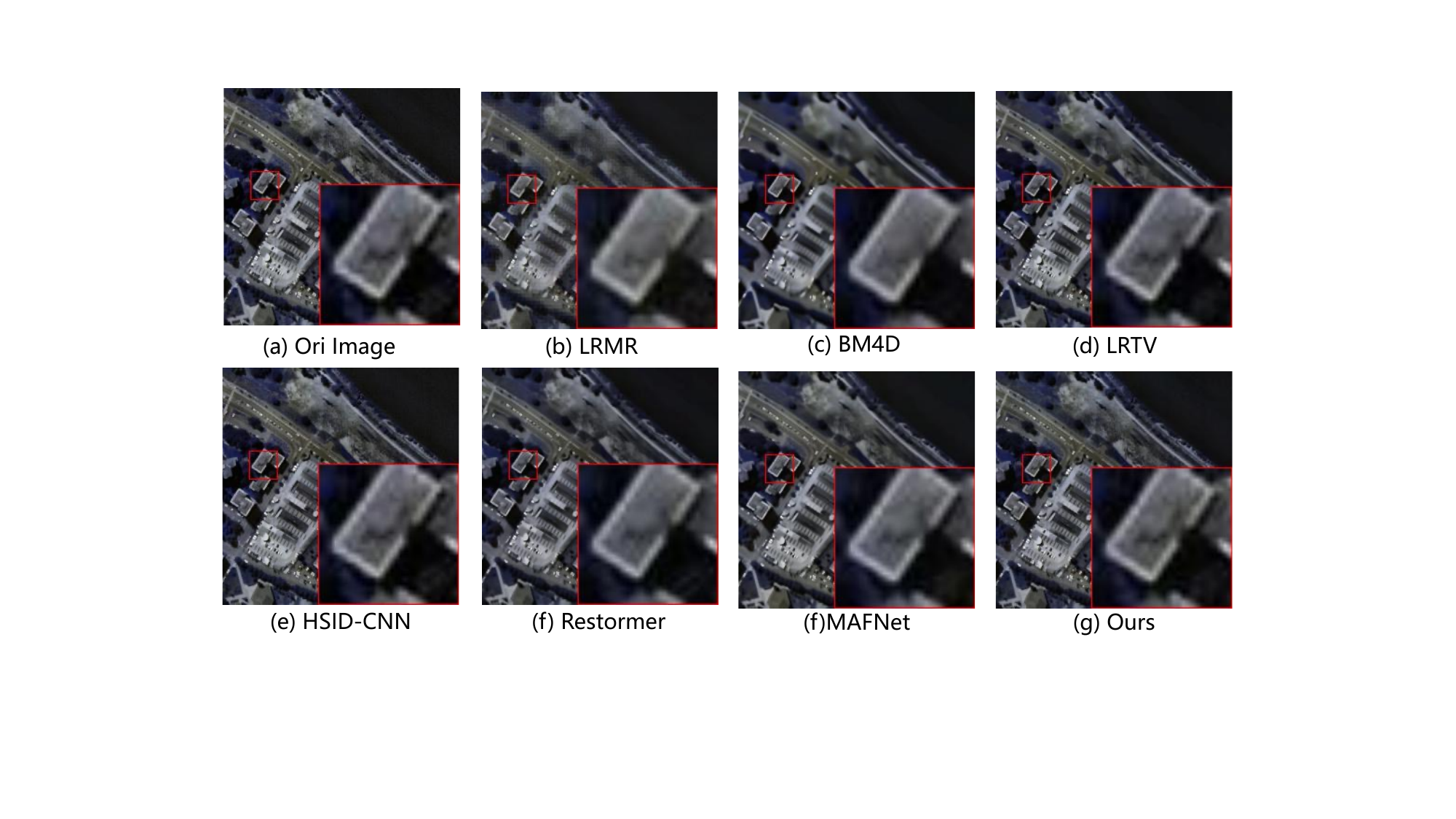}
\caption{Real noise removal results on Pavia dataset with bands (17, 20, 30).}
\label{result2}
\end{figure}

Table \ref{tab1} showcases the results of all methods in different intensities of Gaussian noise. From this table, it is easy to find that our HCANet outperforms all other methods in terms of all metrics when the data contains single-type noise. Additionally, Table \ref{tab2} presents the results with complex noise. We can see that our HCANet achieves the best performance in all noise cases. In addition to quantitative analysis, we also conducted qualitative comparisons, as illustrated in Figs. \ref{result1} and \ref{result2}. False-color denoising results were obtained from three bands (17, 20, 30). It is clear that traditional denoising methods struggle with complex noise. While MAFNet removes most noise, it tends to oversmooth, losing some image details. Restormer performs poorly on complex noise. In contrast, HCANet effectively removes most noise while preserving local details, successfully restoring original image features.

To demonstrate the effectiveness of each component in our HCANet, we conducted ablation study on ICVL dataset. As shown in \ref{tab:ablations}, the HCANet (with all components) performs the best compared with all other variants. Furthermore, we find the performance constantly becomes better as we add local branch, 3D convolution, and MSFN upon the base model. It indicates the necessity of each component in our proposed HCANet.

\begin{table}[]
        \centering
        \caption{\small Ablation studies on ICVL  datasets under noise level $\sigma=30$, epoch = 50}
        \vspace{-3mm}
		\resizebox{\columnwidth}{!}{
			\begin{tabular}{c c c  c c c c  }
				\bottomrule
				Base model & Local branch & 3D Conv  & MSFN & PSNR $\uparrow $& SSIM $\uparrow$ & SAM $\downarrow$ \\
				\midrule
    			\checkmark &  & &  &39.94 &0.965 &0.045 \\
				\checkmark & \checkmark & & &40.58 &0.967 &0.039 \\
				\checkmark  &\checkmark & \checkmark &&42.23 &0.975 &0.037 \\
				\checkmark  &\checkmark & \checkmark &\checkmark &\bf 42.68 &\bf 0.976 & \bf 0.035 \\
				\bottomrule
	\end{tabular}}\hspace{2mm}\vspace{-1.5mm}
	\label{tab:ablations}\vspace{-2mm}
\end{table}

\begin{table}[htbp]
    \centering
    \footnotesize
    \caption{\small Comparative analysis of computational complexity of models under blind Gaussian noise, epoch = 50}
    \begin{tabular}{c|c|c|c}
        \toprule
        Model & Params & GFLOPS & PSNR \\
        \midrule
        MemNet & {\color[HTML]{4472C4}1.98M} & 32.41 & 35.60 \\
        HSID-CNN & {\color[HTML]{FF0000}0.63M} & 40.80 & 36.72 \\
        MAFNet & 20.17M & 20.15 & {\color[HTML]{4472C4}40.51} \\
        Restormer & 14.88M & {\color[HTML]{FF0000}9.55} & 40.27 \\
        Ours &4.75M & {\color[HTML]{4472C4}11.5} &{\color[HTML]{FF0000}41.21} \\
        \bottomrule
    \end{tabular}
    \label{tab:compare}
\end{table}

We also compared the computational complexity of the models, and the results are shown in Table \ref{tab:compare}. HCANet achieves optimal denoising performance while maintaining a relatively moderate number of parameters and computational complexity.

\section{Conclusion}\label{s4}

In this letter, we propose HCANet, a novel network for HSI denoising. In particular, we proposed the convolution and attention fusion module, CAFM, to fuse both global and local features. Furthermore, we propose the multi-scale feed-forward network, MSFN to extract features from multiple scales and enhances the denoising performance. Experimental results on challenging HSI datasets demonstrate the effectiveness of our proposed model in comparison to the state-of-the-art HSI denoising methods. Our model achieves remarkable denoising performance in terms of both quantitative metrics and visual quality of the reconstructed images.

\bibliography{source}
\bibliographystyle{IEEEtran}

\end{document}